# Oscillation modes of dc microdischarges with parallel-plate geometry

**Ilija Stefanović[1, 4], Thomas Kuschel [1], Nikola Škoro[2], Dragana Marić[2], Zoran Lj Petrović[2] and Jörg Winter[1]**

[1] Institut für Experimentalphysik II, Ruhr-Universität Bochum, 44780 Bochum, Germany

[2] Institute of Physics, POB 68, 11080 Belgrade, Serbia

E-mail: ilija.stefanovic@rub.de

**Abstract.** Two different oscillation modes in microdischarge with parallel-plate geometry has been observed: relaxation oscillations with frequency range between 1.23 and 2.1 kHz and free-running oscillations with 7 kHz frequency. The oscillation modes are induced by increasing power supply voltage or discharge current. For a given power supply voltage, there is a spontaneous transition from one to other oscillation mode and vice versa. Before the transition from relaxation to free-running oscillations, the spontaneous increase of oscillation frequency of relaxation oscillations form 1.3 kHz to 2.1 kHz is measured. Fourier Transform Spectra of relaxation oscillations reveal chaotic behaviour of microdischarge. *Volt-Ampere* characteristics associated with relaxation oscillations describes periodical transition between low current, diffuse discharge and normal glow. However, free-running oscillations appear in subnormal glow only.

## 1. Introduction

Discharge stability is one of the important requirements for the proper functioning of micro-discharge devices. Recently, the attention on stability and different modes of microdischarges were triggered by several studies, like "self-pulsing" microplasma in a microhollow cathode discharge [1] and micro plasma jet instabilities, including analysis of different discharge stages: chaotic, bullet und continuous mode [2].

Earlier investigations on large-scale DC low-pressure discharges have made a large effort to understand the causes of discharge current oscillations and constrictions [3]. In general, oscillations of gas discharges may be associated with a combination of the physical processes occurring in the discharge itself and the external circuit. For example in the low current diffuse regime (Townsend's dark discharges) space charge effects lead to a negative differential resistance [4, 5] which supports oscillations if the effective round the loop resistance is negative [4]. Phenomenological descriptions often involved assigning effective circuit elements and their values to the discharge [6,7]. Physical models identify properties of charged particles and their distribution [4, 8] as causing instabilities.

Numerous regimes of oscillations may be identified [4,9]. Some oscillations may be of the relaxation type whereby the discharge is repeatedly turning itself off through the combined effect of increased current and external circuit. Others may be associated with transitions between different regimes of operation or spatial modes and some undulations of voltage and current may be due to variable properties on the surface when the discharge is moving [10]. Discharge current oscillations are often associated with complex instabilities and even chaotic behaviour has been invoked to explain the phenomena [11-14]. As for the micro discharges it has been noticed that a number of different spatial modes may exist and discharges may be localized over a small area of the electrodes since the diffusion length at corresponding pressures is quite small [15]. As a result, transitions between different spatial modes may cause instabilities. A paper by Maguire et al [15] deals with such instabilities and oscillations in low current breakdown in micro discharges. Here we extend the analysis of the standard size discharges to the smaller scale, higher-pressure discharges with an aim to understand the basic processes that influence the stability of micro-plasmas.

---

[4]Author to whom any correspondence should be addressed

## 2. Experiment

The electrode system consists of highly polished parallel-plate stainless steel electrodes with variable distance controlled by a linear stage (see figure 1). The electrode diameter is 5 mm and the distance was set to $d = 1$ mm. The working gas is argon at pressure $p = 10$ Torr so that $pd$ is 10 Torr mm (1 Torr cm), at which point the Paschen curve for Argon has a minimum breakdown voltage of about 220 V. Precision valves maintain the continuous gas flow. The flow is not measured but it is kept as small as possible to avoid perturbations of the discharge. The electrodes are sealed with Teflon and tightly fitted in to the vacuum-sealed Plexiglas chamber to avoid long-path breakdown [16]. High DC negative voltage ($U$) is applied to the cathode to ignite the discharge and the discharge current is controlled by changing voltage and/or by changing the loading resistance ($R_s$), as described earlier [3]. The current monitor resistance ($R_m$) is used to measure the discharge current. The discharge voltage $U_{CA}$ is the difference between cathode potential $V_c$ and anode potential $U_m$, $U_{AC} = V_C - V_m$. Voltage and current waveforms are simultaneously measured by means of digital oscilloscope and archived in the computer for latter evaluation.

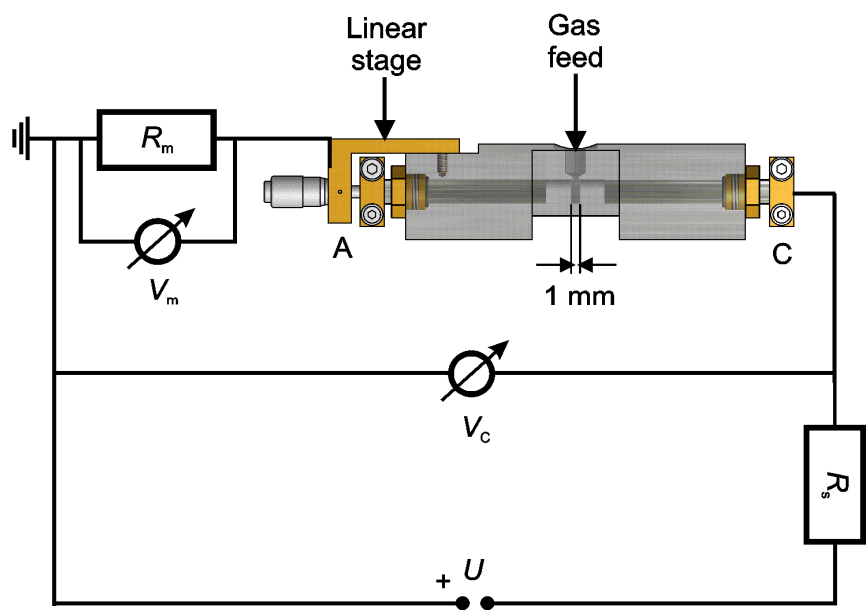

**Figure 1**. Experimental set-up: A - anode, C- cathode, $R_S$ - loading resistance and, $R_m$ - current monitor resistance, $U$ - power supply voltage, $V_c$ - cathode potential in respect to the ground, $V_m$- anode potential in respect to the ground.

## 3. Results

### 3.1. Different oscillation modes

Typical voltage (upper waveform) and current (lower waveform) waveforms displaying oscillations are presented in Figure 2. Two different oscillating modes can be recognized: low frequency relaxation oscillations with increasing frequency ($f_r$) (the increase in frequency may be difficult to observe) and high frequency free-running oscillations with an increasing amplitude ($f_f$).

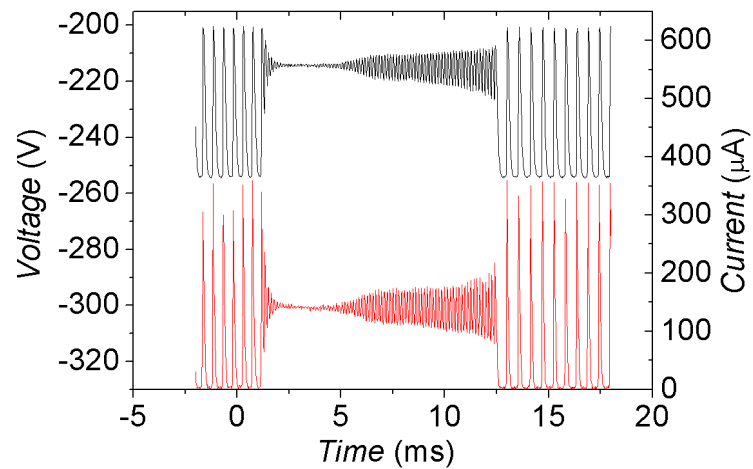

**Figure 2**. Voltage black line) and current (red line) oscillations in parallel plate microplasma with continuous argon flow. Discharge parameters are: pressure $p$ = 10×Torr, electrode distance $d$ = 1mm.

The low frequency oscillations are relaxation oscillations, during which the discharge periodically ignites and chokes [3]. From the figure 2 one can see that current changes form 350 µA to almost zero and vice versa. After a closer observation we find that the discharge switches from a high current to a low-current regime with the current density of about 5 µA/cm$^2$, following the time resolved development of the low-pressure discharges (see [17]). In the low-current regime, the discharge is running either as self-sustained or non-self-sustained Townsend discharge. The high-current regime is either normal or abnormal glow, depending on the current value. During the transition from low current to the high-current regime, the discharge runs through different modes: subnormal glow, constricted (normal) and abnormal glow, depending on the *pd* value [17].

In high frequency oscillatory mode (which starts from 1.2 ms and finishes at 12.5 ms on figure 2), the discharges run in three different stages: (1) damped oscillations (1.2 ms - 2.5 ms), (2) free-running oscillations with about 7 kHz frequency (4 ms - 12.5 ms), and (3) in quasi-stationary regime (2.5 ms - 4 ms). These oscillations are similar to oscillations observed in large-scale discharges, with electrode separation of 1 -3 cm and the discharge pressure 0.5 - 2 Torr (for details see ref. [3-5]). Soon after transition from relaxation oscillations to high-frequency oscillations (at approximately 1.2 ms, figure 2), oscillations vanish due to the strong damping. In high-frequency mode the current and voltage oscillation amplitude is not as high as for relaxation oscillations and the discharge does not switch off but runs with voltage and current oscillating around some mean value. At 4 ms, the oscillations appear again and start to grow in amplitude. The mean current is varying from 145 µA to 137 µA, when the discharge again switches from the low frequency to high-frequency relaxation oscillations. Afterwards, the whole process repeats.

Different oscillation modes and their combinations can be observed/controlled by changing the power supply voltage (figure 3 (a) – figure 3 (f)). Discharge switches "on" and "off" periodically (figure 3a) at the maximum power supply voltage. During the "on" phase (60 ms) the discharge current drops from 180 µA to 130 µA but the voltage stays almost constant, indicating that the discharge runs in the normal glow. The discharge spontaneously switches "off" for the next 30 ms and then switches "on" again.

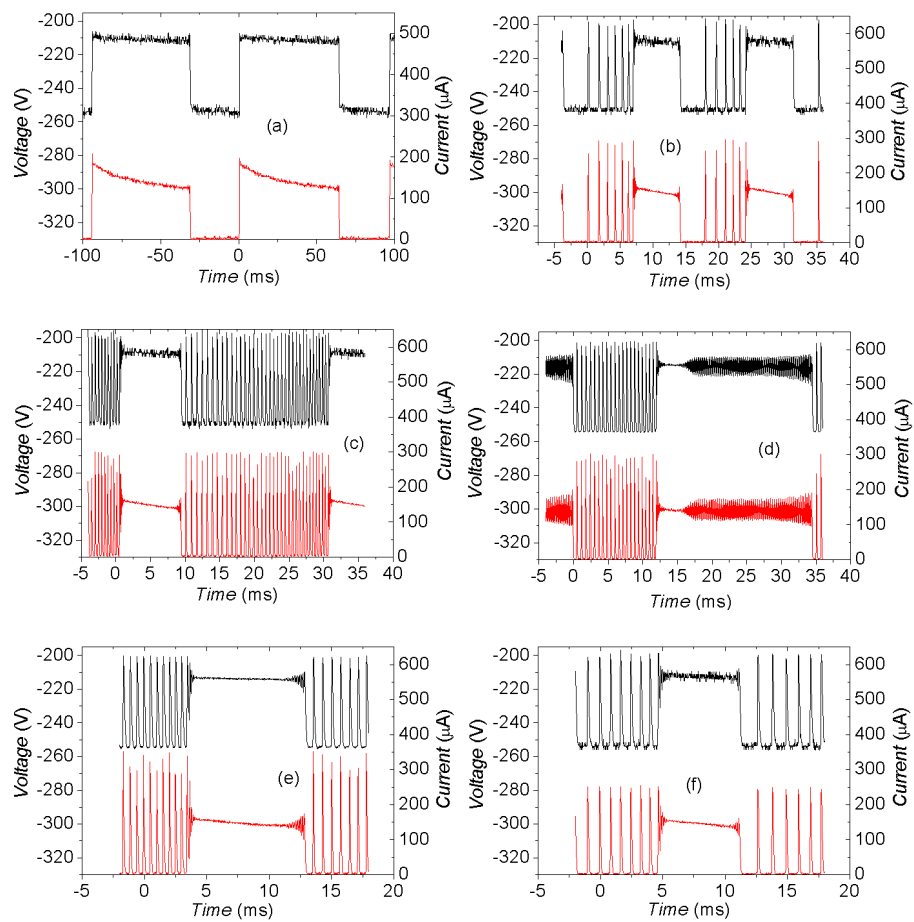

**Figure 3 (a) – (f).** Voltage (black line) and current (red line) oscillations in parallel plate microplasma with continuous argon flow for different power supply voltages: (a) the highest voltage, (f) the lowest voltage. Discharge parameters are: pressure $p = 10$ Torr, electrode distance $d = 1$ mm, electrode radius $r = 2.5$ mm.

By decreasing the power supply voltage ($U$ in figure 1) the relaxation oscillations appear, first with increasing and then decreasing frequency, thus replacing effectively the "off" period.

Sometimes, the relaxation frequency increases spontaneously, without changing the power supply voltage $U$. This effect is connected to the chaotic behaviour of the discharge that shall be discussed later. The low frequency oscillations end spontaneously and the damped high-frequency oscillations start associated with the transition to the quasi-stationary discharge period. Soon after the high frequency oscillations disappear due to high damping. At the same time the quasi dc current drops -continuously from 150 µA to 135 µA. At the end of this phase the high-frequency oscillations spontaneously start again and grow in the amplitude until the discharge switches to the low-frequency oscillation mode (figure 3b). A further decrease in the power supply voltage $U$ leads to an increase of the frequency of relaxation oscillations (the low frequency mode) (figure 3 (c)). The relaxation oscillation frequency spontaneously increases from 1.3 kHz - 2.1 kHz (figure 3 (c)). Afterwards, the discharge changes to the high-frequency mode with strongly damped oscillations. Further decrease of the power supply voltage (figure 3 (d)) leads to the more developed high-frequency oscillations of 7 kHz frequency. After the quasi-stationary pulse has been established, those oscillations appear and grow in amplitude for a while, but then they stay stable for more then 15 ms. Before the transition to the relaxation oscillations the high frequency amplitude starts growing again. The frequency of relaxation oscillations remains constant at about 1.7 kHz. It appears that here we have caught a possible situation where a period of growth of high frequency oscillations superimposed on the quasi dc current and voltage, remained stable for a while in other cases it leads rapidly to the development of the relaxation oscillations.

By the next decrease of the power supply voltage the high-frequencies are suppressed (very rapidly damped) most of the cycle and the discharge runs in quasi-steady-state regime with a constant voltage and a slightly decreasing current from 160 µA to 140 µA (figure 3 (e)). The relaxation oscillations are changing their frequency again starting from 1.4 kHz to 1.9 kHz before the discharge switches to the high-frequency regime. Figure 3 (f) shows the lowest power supply voltage used. The discharge again switches spontaneously from one to other discharge mode: low frequency changes from 1 kHz to 1.3 kHz and the high frequency is constant at $f = 7$ kHz but strongly suppressed compared to the high-frequency oscillations in figure 3 (d).

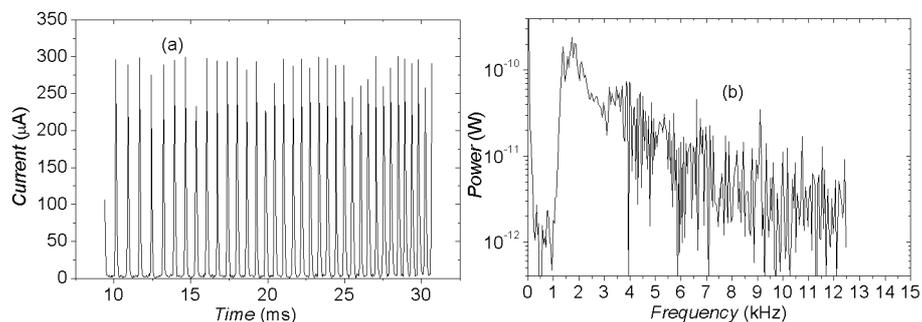

**Figure 4.** (a) The portion of current waveform from figure 3c showing the relaxation oscillations with increasing frequency; (b) Fourier-transform power spectrum of relaxation oscillations.

The most exciting observation regarding the low-frequency relaxation oscillations is that their frequency increases spontaneously. To analyse this closely we focus on the discharge relaxation oscillations from figure 3c between 9.4 ms to 30.7 ms. This is presented in figure 4(a) and the power spectrum of the Fourier-transform of this data is presented in the figure 4(b). The power spectrum of current relaxation oscillations (figure 4(b)) shows a broad distribution characteristic of chaotic behaviour of dissipative dynamical systems. Similar nonlinear dynamics was previously observed for thermionic discharges with volume ionisation [12], double plasma devices [13] and recently in cold atmospheric pressure plasma jet [2]. However, to analyse the character of the non-linear dynamical system properly, the phase-space reconstruction of the dynamical space is needed [18]. The oscillation modes and their behaviours depend on the current density and on the other discharge parameters, such *E/n*, and they should be investigated further.

The driving force for transition from high frequency to the low-frequency oscillations could be the gas heating in the active discharge zone, where the power dissipation is $P = 3.5*10^6$ W/m$^3$. (For a comparison, the power density of a 60 W fluorescent lamp is about an order of magnitude smaller). As the temperature increases due to a high power deposition, the local pressure of Argon in the microdischarge is reduced and thus the *E/n* increases. If conditions are right, this could drive the discharge to the low-frequency relaxation oscillation mode.

Phelps et al [3,4] have identified causes for low current (Townsend regime) oscillations at standard sizes (*d* ~ 1 cm) /pressures (*p* ~ 1 Torr). They found it to be the negative differential resistance caused by the space charge modification of the field close to the cathode and its effect on the secondary electron yield. Combination of the negative differential resistance and external circuit that leads to overall negative resistance in the circuit loop will lead to free running oscillations. The other issue would be the possibility that while changing the active spot the properties at the surface are such that the discharge may change its current dramatically. Thus, transitions between different spatial modes may lead to oscillations. Mahoney et al. [15] have already observed numerous modes of oscillations in micro discharges. It is also worth noting that it has been found that even tightly fitting microdischarges have high enough pressure to allow long path breakdown around the edges of electrodes [16,19]. Certainly, both standard explanations of the causes of oscillation and the effects that become important at higher pressures such as heating (and depletion of atoms leading to a higher *E/n*) may play a role and it would be of interest to do detail modelling of this issue.

What drives the transition from low frequency to the high frequency is difficult to explain exactly at the moment.

*3.2 V-A characteristics*

We analyse the discharge *V-A* characteristics in order to further understand the conditions associated with oscillations. The analyses of the time series of two different observables (voltage and current) may help to understand better the properties of the discharge as a dynamical system. As a difference to the *static V-A* characteristic, which describes the discharge's voltage and current in steady state, the *dynamic V-A* characteristic describes the discharge transient behaviour and has usually hysteresis. The dynamic *V-A* characteristics of the microdischarge are presented in figure 5(a)-(f). These figures correspond each to the current and voltage waveforms presented in figure 3(a)-(f).

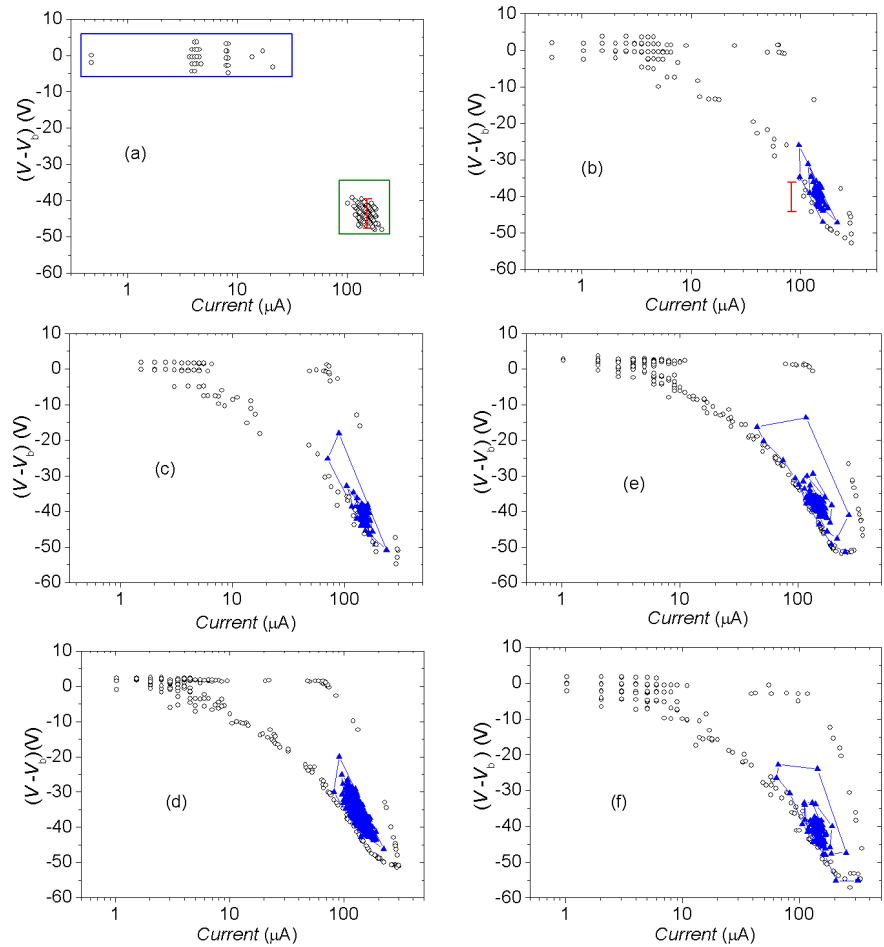

**Figure 5 (a) – (f).** Discharge *V-A* characteristics extracted from current and voltage waveforms. Figure 5a two rectangular forms present the low-current (blue) diffuse mode and normal glow mode (green). Opened circles: low-frequency relaxation oscillations; blue triangles: high-frequency oscillations. Discharge voltage is normalized to the breakdown voltage ($V_b$) to stress the voltage change with increasing current. The vertical red solid lines on figure (a) and (b) shows voltage error bar.

Figure 5(a) presents the discharge running in two different modes: low current, diffuse (Townsend-like) discharge, with currents ranging from 0.4 µA to 20 µA, and high-current normal glow mode. In the normal glow mode, during 65 ms, the discharge current decreases from 185 µA to 125 µA and the discharge voltage stays almost constant, considering the large voltage error bars (about ±5 V). The microdischarge switches spontaneously from one mode to other, staying in each mode for few tens of milliseconds.

When oscillations are present, two different oscillations modes can be distinguished: low frequency relaxation oscillations and high frequency oscillations. The characteristic hysteresis, previously reported for hollow-cathode discharges [19, 20] and parallel-plate discharges [17,22] for the relaxation oscillations is present: the discharge current starts from low current, diffuse regime (1 µA) and runs through the upper branch to the maximum current value (about 300 µA), which characterize the glow discharge. The upper branch characterises the capacitive nature of the discharge: current increase advances the voltage drop. After reaching their maximum values, the voltage changes by a few volts while the current decreases from 300 µA to 100 µA, as for the normal glow. The $j/p^2$ value is similar to the normal glow region of the large-scale discharges [23]. Measurements of axial discharge profiles [24] showed that for the relaxation oscillations the discharge reaches even the abnormal glow with the negative glow closer to the cathode then in the normal glow. Afterwards, the current and voltage drop and the discharge turns back to the Townsend mode through the lower brunch of the *V-A* characteristics (solid triangles on figure 5b-f).

At some point, the discharge spontaneously switches to the high-frequency mode, without any change in outer circuitry. The figure 5(b) -(f) presents high-frequency oscillations with solid triangles connected with the solid line. The amplitude of oscillations first exponentially decays to the quasi-steady-state value, when the dynamic *V-A* characteristics meets static *V-A* characteristics. Depending on discharge conditions (in this case different power supply voltage) the discharge stays longer (figure 3(e)) or shorter (figure 3(d)) in the quasi-steady state, with slightly decreasing current. High-frequency oscillations start again and then grow in amplitude until reaching the point when they spontaneously switch to the relaxation oscillations (see figure 5 (b)-(f)).

## 4. Conclusion

Different oscillation modes of parallel-plane microdischarge in argon were recorded and presented. The discharge instabilities appear to be a consequence of the negative slope of *V-A* in the transition from Townsend-like, diffuse discharge to normal or abnormal glow discharge, where the space-charge effects become dominant. Two different instability modes were detected: low frequency and high frequency oscillations. Low frequency oscillations are relaxation oscillations, where the discharge periodically switches between low-current and high-current modes. These are similar to the self-pulsing oscillations of micro-hollow cathode discharges. In high frequency discharge mode the discharge current and voltage oscillate with smaller amplitude and approach the point of static *V-A* characteristics. Switching from one to other oscillation mode is spontaneous. In some cases, before the transition from low- to high-frequency mode, the oscillation frequency increases and the Fourier power spectrums has a form of dynamical system with chaotic behaviour. The stability and dynamical properties of the microdischarges are a very important issue, thus, as the mechanism for frequency increase and chaotic behaviour cannot be pinpointed clearly, it should be further investigated.

**Acknowledgment**
This work was supported by DFG FOR 1123, Research Department "Plasmas with Complex Interactions, DAAD Grant 50430276 and by MNTRS 171037 project.